\documentclass[pdflatex,sn-mathphys]{sn-jnl}
\usepackage{graphicx}
\usepackage{hyperref}
\usepackage{pifont}
\usepackage{threeparttable}
\usepackage{makecell}
\usepackage{multicol}
\usepackage{multirow}
\usepackage{booktabs}
\usepackage{makecell}
\usepackage{adjustbox}
\usepackage{float}
\usepackage{bm}
\usepackage{xcolor} 

\jyear{2023}%

\theoremstyle{thmstyleone}%
\theoremstyle{thmstyletwo}%

\theoremstyle{thmstylethree}%

\begin{document}

\title[LEIGNN]{Lightweight equivariant model for efficient machine learning interatomic potentials}

\author[1,2]{\fnm{Ziduo} \sur{Yang}}\email{yangzd@mail2.sysu.edu.cn}
\equalcont

\author[3]{\fnm{Xian} \sur{Wang}}\email{sunnywx623@163.com}
\equalcont{These authors contributed equally to this work.}

\author[1]{\fnm{Yifan} \sur{Li}}\email{e0576095@u.nus.edu}

\author[1,2]{\fnm{Qiujie} \sur{Lv}}\email{lvqj5@mail2.sysu.edu.cn}

\author*[4,5,6]{\fnm{Calvin Yu-Chian} \sur{Chen}}\email{cy@pku.edu.cn}

\author*[1,7]{\fnm{Lei} \sur{Shen}}\email{shenlei@nus.edu.sg}

\affil[1]{\orgdiv{Department of Mechanical Engineering}, \orgname{National University of Singapore}, \orgaddress{\city{Singapore}, \postcode{117575}, \country{Singapore}}}

\affil[2]{\orgdiv{Artificial Intelligence Medical Research Center, School of Intelligent Systems Engineering}, \orgname{Shenzhen Campus of Sun Yat-sen University}, \orgaddress{\city{Shenzhen}, \postcode{518107}, \country{China}}}

\affil[3]{\orgdiv{Department of Physics}, \orgname{National University of Singapore}, \orgaddress{\city{Singapore}, \postcode{117551}, \country{Singapore}}}

\affil[4]{\orgdiv{AI for Science (AI4S)-Preferred Program, School of Electronic and Computer Engineering}, \orgname{Peking University Shenzhen Graduate School}, \orgaddress{\city{Shenzhen}, \postcode{518055}, \country{China}}}

\affil[5]{\orgdiv{State Key Laboratory of Chemical Oncogenomics, School of Chemical Biology and Biotechnology}, \orgname{Peking University Shenzhen Graduate School}, \orgaddress{\city{Shenzhen}, \postcode{518055}, \country{China}}}

\affil[6]{\orgdiv{Guangdong L-Med Biotechnology Co., Ltd}, \orgname{Meizhou}, \orgaddress{\city{Guangdong}, \postcode{514699}, \country{China}}}

\affil[7]{\orgdiv{National University of Singapore (Chongqing) Research Institute}, \orgaddress{\city{Chongqing}, \postcode{401123}, \country{China}}}


\abstract{In modern computational materials science, deep learning has shown the capability to predict interatomic potentials, thereby supporting and accelerating conventional simulations. However, existing models typically sacrifice either accuracy or efficiency. Moreover, lightweight models are highly demanded for offering simulating systems on a considerably larger scale at reduced computational costs. Here, we introduce a lightweight equivariant interaction graph neural network (LEIGNN) that can enable accurate and efficient interatomic potential and force predictions for molecules and crystals. Rather than relying on higher-order representations, LEIGNN employs a scalar-vector dual representation to encode equivariant features. By learning geometric symmetry information, our model remains lightweight while ensuring prediction accuracy and robustness through the equivariance. Our results show that LEIGNN consistently outperforms the prediction performance of the representative baselines and achieves significant efficiency across diverse datasets, which include catalysts, molecules, and organic isomers. Furthermore, we conduct molecular dynamics (MD) simulations using the LEIGNN force field across solid, liquid, and gas systems.  It is found that LEIGNN can achieve the accuracy of \textit{ab initio} MD across all examined systems.}

\maketitle

\section{Introduction}
In the field of computational materials, calculating interatomic potentials is critical for obtaining energy and related physical quantities such as forces and atomic trajectories. Computational methods that use pre-fitted empirical functions to form interatomic potentials, such as classical molecular dynamics (MD), provide very fast but less accurate material-property calculations. Meanwhile, methods based on high-fidelity quantum-mechanics calculations, such as density functional theory (DFT) and \textit{ab initio} molecular dynamics (AIMD), offer highly accurate energies and forces but require high computational costs.

\begin{figure*}[htp]
  \centering
  \includegraphics[width=11.8cm]{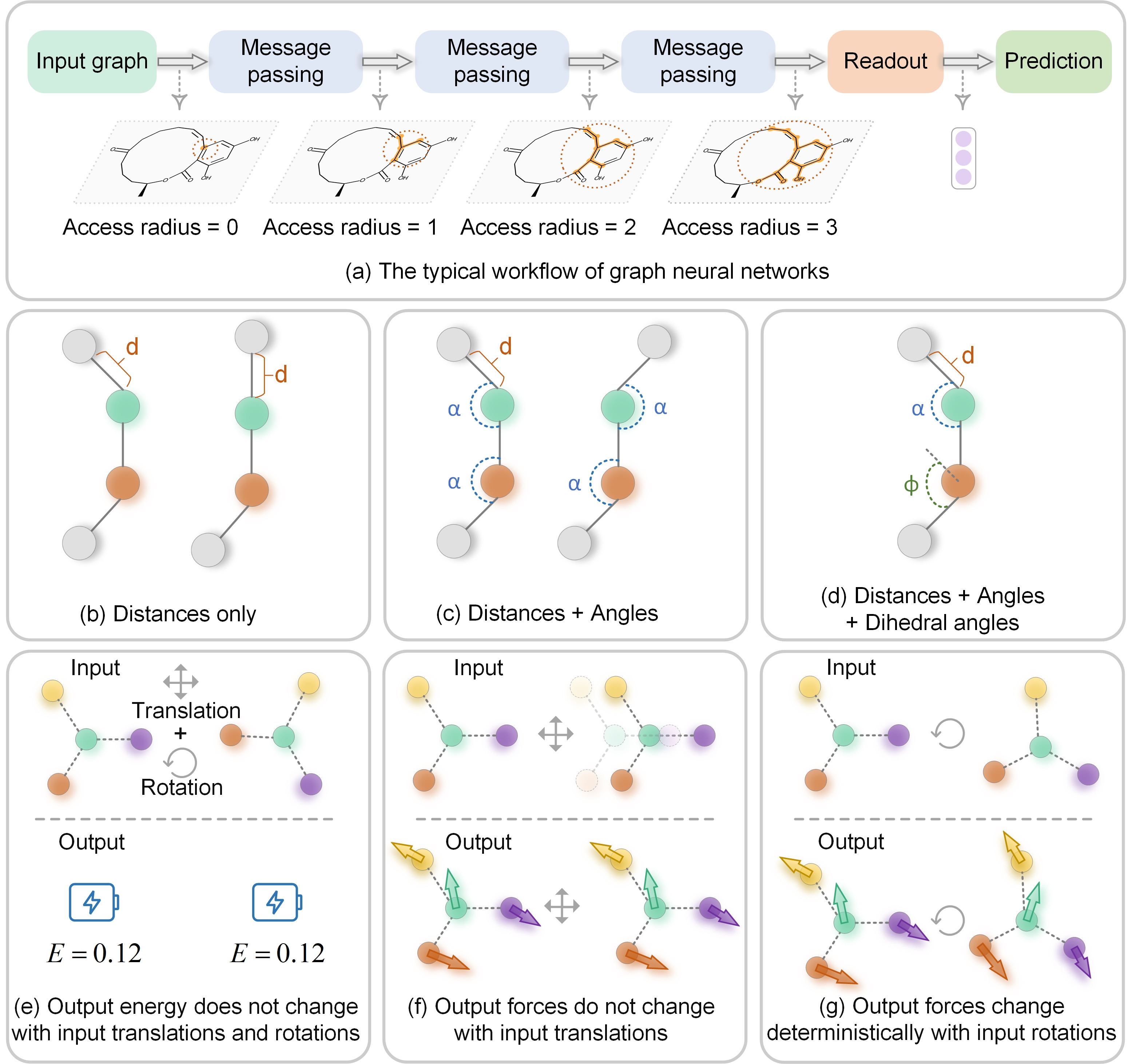}
  \caption{Overview of GNNs applied to materials property prediction. (a) A typical GNN workflow is illustrated, emphasizing how GNNs can include a wider range of interactions (or message passing) by stacking multiple layers to extend the accessible radius. (b), (c), and (d) underscore the importance of integrating structural features into GNNs, which include: (b) distance only, (c) both distance and angles, and (d) all distance, angles, and dihedral angles. (e), (f), and (g) elucidate the concepts of invariance and equivariance within the context of energy and forces prediction.}
  \label{fgr:GNNs}
\end{figure*}

To address the above dilemma, deep learning techniques, such as graph neural networks (GNNs), have been proposed for predicting interatomic potentials \cite{batatia2022mace, batzner20223, chen2022universal, xie2024GPTFF} or DFT Hamiltonian \cite{zhong2023transferable, zhong2024universal, li2022deep, wang2024deeph} in speed while preserving quantum mechanics-level accuracy. \textcolor{black}{These models can be mainly divided into two categories: invariant models \cite{Zhang2018, Yoon2020, deng2023chgnet, pablo2023fast, li2024jmi, dai2021graph, hu2021forcenet}, such as CGCNN \cite{xie2018crystal}, SchNet \cite{schutt2017schnet}, MEGNet \cite{chen2019graph}, ALIGNN \cite{choudhary2021atomistic}, and M3GNet \cite{chen2022universal}, and equivariant models \cite{han2024survey, gong2023general, fuchs2020se, brandstetter2021geometric, satorras2021n}, such as MACE \cite{batatia2022mace}, NequIP \cite{batzner20223}, ScN \cite{zitnick_scn_2022}, eScN \cite{passaro2023reducing}, and DeepRelax \cite{yang2024deeprelax}.} To represent the molecule or crystal structure, GNNs typically use a graph where nodes are atoms and edges are bonds between atoms. Atomic interactions are then simulated by applying graph convolution operations on the graph, where an atom can access its neighboring atoms during this process (\textbf{Fig. \ref{fgr:GNNs}a}).

Currently, the most widely used GNNs are designed with architectures that are invariant to transformations from the Euclidean group E(3), ensuring consistent output for energy relative to translations, rotations, and reflections. This is achieved by leveraging invariant features, such as bond lengths and angles, which remain constant under these transformations. Early models like CGCNN \cite{xie2018crystal}, SchNet \cite{schutt2017schnet}, and MEGNet \cite{chen2019graph} primarily incorporate bond lengths, leading to challenges in distinguishing structures with identical bond lengths but different overall configurations (\textbf{Fig. \ref{fgr:GNNs}b}). Later iterations, like DimeNet \cite{gasteiger_dimenet_2020}, ALIGNN \cite{choudhary2021atomistic}, and M3GNet \cite{chen2022universal}, improve upon this by integrating bond angles. Despite this improvement, they still struggled to differentiate between structures sharing the same angles (\textbf{Fig. \ref{fgr:GNNs}c}). Recent models, such as GemNet \cite{gasteiger2021gemnet} and SphereNet \cite{liu2022spherical}, propose considering dihedral angles in GNNs to unambiguously recognize the local structures (\textbf{Fig. \ref{fgr:GNNs}d}). It is worth noting that distance and angular features are invariant representations which are only used to keep the geometric symmetry of crystals with respect to E(3) transformations, rather than utilize the geometric symmetries in a more profound manner for increasing the prediction accuracy and the sample efficiency. Actually, the idea to leverage crystal symmetry for effectively describing the electron wave function and material properties was proposed a century ago. Felix Bloch demonstrated a translation-symmetry-based structural function to maintain the equivariance of wave functions in 1928. Such an equivariant idea in crystals with periodic structures in 3D space provides a powerful framework for accurately understanding the material properties and significantly reduces the cost of computation, opening a new era for computational material science. \textbf{Fig. \ref{fgr:GNNs}e-g} offer a concise elucidation of invariance and equivariance in the context of predicting energy and forces.

In this work, we integrate equivariance into GNNs for actively exploiting crystal symmetries and offering a richer geometrical representation compared to their invariant counterparts. Most importantly, our equivariant network is lightweight, incorporating only scalar and vector features in a manner that preserves symmetry. Our lightweight equivariant interaction graph neural network (LEIGNN) is different from previously equivariant GNNs which are based on high-order functions, such as the spherical harmonic function. While these high-order functions have been associated with increased accuracy compared to invariant GNNs, they typically come at the cost of increased computational expense \cite{atz2021geometric, brandstetter2021geometric, fuchs2020se, satorras2021n, shuaibi2021rotation, batzner2023advancing, musaelian2023learning, batzner20223, passaro2023reducing, zitnick_scn_2022, batatia2023equivariant, batatia2022mace}. For example, cutting-edge models like NequIP \cite{batzner20223}, MACE \cite{batatia2022mace}, ScN \cite{zitnick_scn_2022}, and eScN \cite{passaro2023reducing} employ tensor product operations to combine input features and filters in an equivariant manner, which is computationally demanding in practice \cite{fu2023forces, zitnick_scn_2022, passaro2023reducing}.

Our model aims to achieve both accurate and efficient predictions for interatomic potentials and forces. Specifically, we assign each node both scalars and vectors to represent equivariant features. LEIGNN combines these entities in a symmetry-preserving fashion to maintain equivariance. Equivariant LEIGNN surpasses scalar-only invariant models \cite{xie2018crystal, schutt2017schnet, chen2019graph} in accuracy and generalization ability. It also offers a lightweight structure compared to high-order tensor models \cite{batzner20223, zitnick_scn_2022, batatia2022mace}. We also conduct MD simulations using LEIGNN potential across solid, liquid, and gas systems. LEIGNN can achieve the \textit{ab initio} MD accuracy with highly computational efficiency across all examined systems, showing its accuracy and efficiency.

\begin{figure}[htp]
  \centering
  \includegraphics[width=11.8cm]{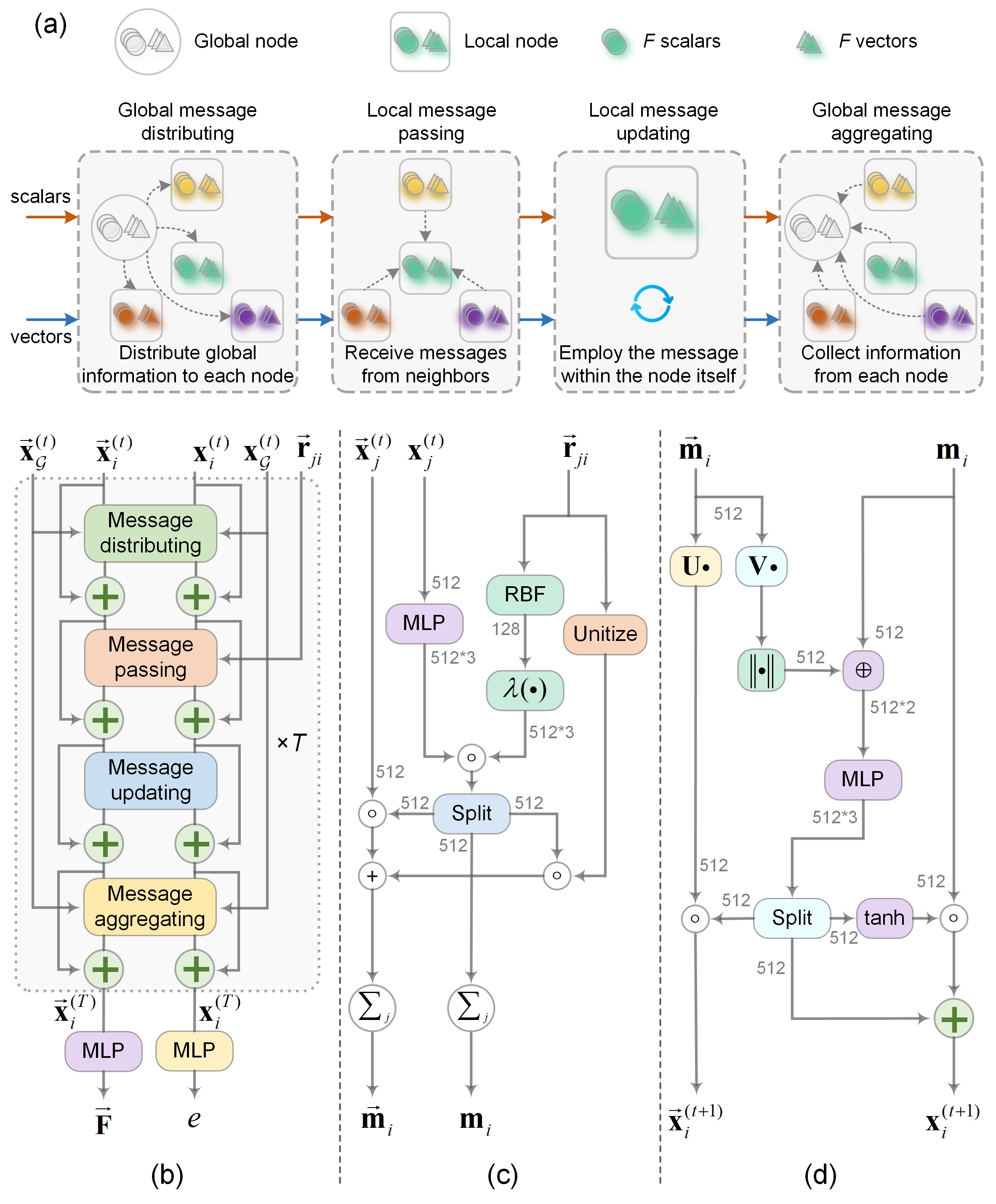}
  \caption{{\color{black}The overall architecture of LEIGNN. (a) LEIGNN uses message distributing, message passing, message updating, and message aggregating to iteratively update node representations. (b) LEIGNN consists of $T$ layers. (c) The message passing phase. (d) The message updating phase. The number of features after each operation is annotated in grey.}}
  \label{fgr:LEIGNN}
\end{figure}

\section{Methods}

\subsection{Network achitecture}
\textcolor{black}{The proposed LEIGNN aims to enhance GNNs by incorporating equivariance, offering a richer geometric representation while retaining a lightweight model. Each node in LEIGNN is assigned scalars and vectors to represent invariant and equivariant features, respectively. This approach is similar to PaiNN \cite{schutt2021equivariant} and NewtonNet \cite{haghighatlari2022newtonnet}, but with a more powerful message update scheme and global interaction modeling.}

\textcolor{black}{LEIGNN gradually updates the node representations through four key processes: global message distributing, local message passing, local message updating, and global message aggregating, as illustrated in \textbf{Fig. \ref{fgr:LEIGNN}a} and detailed in \textbf{Fig. \ref{fgr:LEIGNN}b-d}. The local message passing phase aggregates information from neighboring nodes to simulate two-body interactions. The local message updating phase integrates $F$ scalars and $F$ vectors within a node to update node representations. The global message aggregating and distributing phases promote long-range information exchange by maintaining a global scalar and vector.}

\subsection{Problem definition}
In this work, the atomic structure is represented as a 3D interaction graph $\mathcal{G} = (\mathcal{V}, \mathcal{E}, \mathcal{R})$, where $\mathcal{V}$ and $\mathcal{E}$ are sets of nodes and edges. $\mathcal{R}$ are sets of 3D coordinates of nodes. Each node is connected to its closest neighbors within a cutoff distance $D$ with a maximum number of neighbors $N$, where $D$ and $N$ are predefined numbers. Each node $v_i \in \mathcal{V}$ has its scalar feature $\mathbf{x}_i \in \mathbb{R}^F$, vector feature $\vec{\mathbf{x}}_i \in \mathbb{R}^{F \times 3}$ (i.e., retaining $F$ scalars and $F$ vectors for each node), and 3D coordinate $\vec{\mathbf{r}}_i \in \mathbb{R}^3$. The scalar and vector features can be updated during training. We set the number of features $F$ as a constant throughout the network. The scalar feature is initialized to an embedding only dependent on the atomic number as $\mathbf{x}_i^{(0)}=E(z_i) \in \mathbb{R}^F$, where $z_i$ is the atomic number and $E$ is an embedding layer that takes $z_i$ as input and returns a $F$-dimensional feature. This embedding is similar to the one-hot vector but is trainable. The vector feature is set to $\vec{\mathbf{x}}_i^{(0)}=\vec{\mathbf{0}} \in \mathbb{R}^{F \times 3}$ at the initial step. We also define $\vec{\mathbf{r}}_{ij}=\vec{\mathbf{r}}_j-\vec{\mathbf{r}}_i$ as a vector from node $v_i$ to node $v_j$. Norm $\Vert \cdot \Vert$ and dot product $\cdot$ are calculated along the spatial dimension, while concatenation $\oplus$ and Hadamard product $\circ$ are calculated along the feature dimension. Given a set of 3D interaction graphs, our goal is to learn a model $f$ to predict the potentials and forces as $f(\mathcal{G})=(e, \vec{\mathbf{F}})$, where $e \in \mathbb{R}^{1}$, $\vec{\mathbf{F}} \in \mathbb{R}^{M \times 3}$ and $M$ is the number of nodes in $\mathcal{G}$.

\subsection{Local message passing}
\begin{figure}[htbp]
  \centering
  \includegraphics[width=11.8cm]{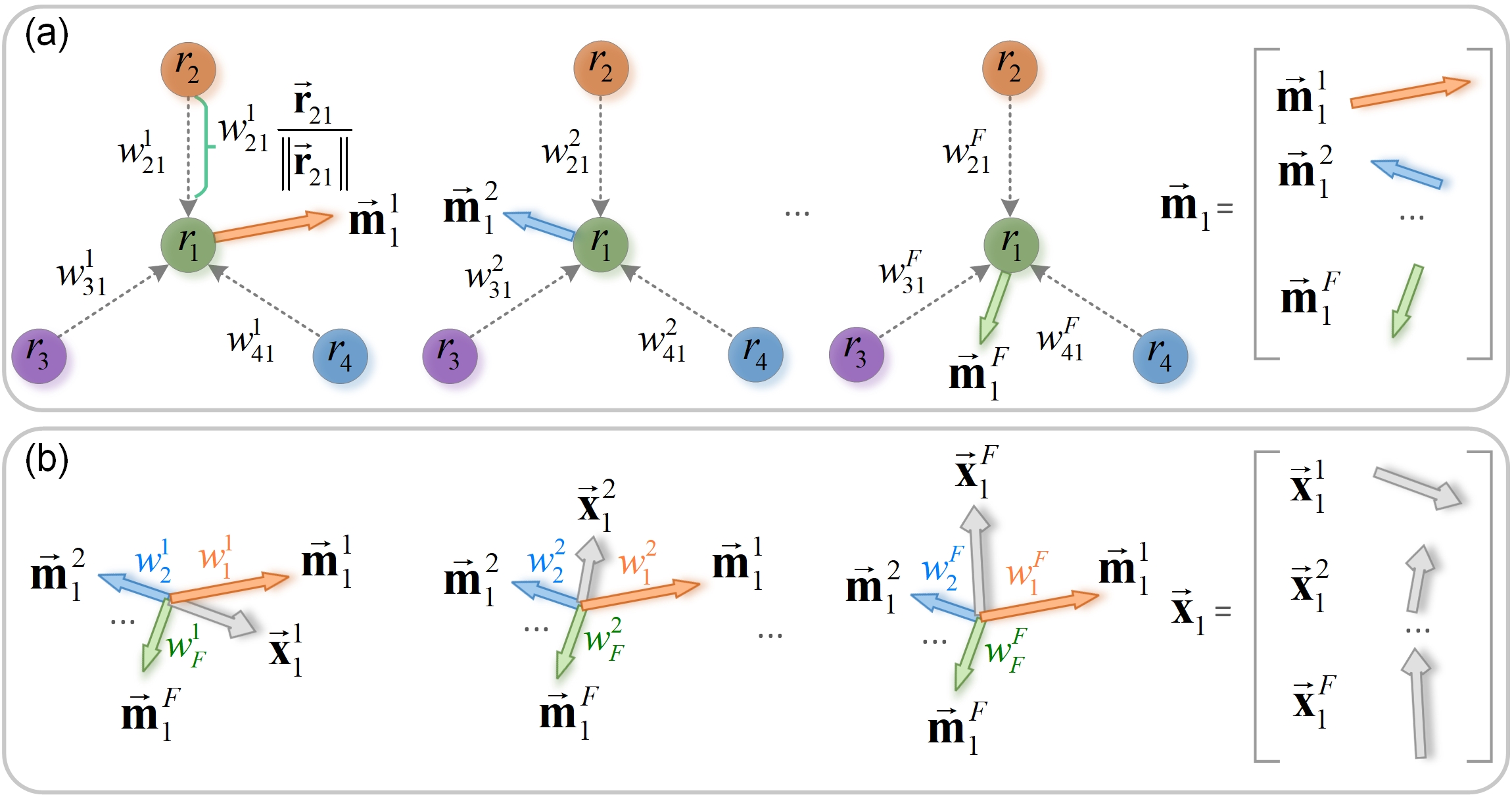}
  \caption{{\color{black}An illustration of how vector representation evolves during message passing phase and message updating phase. (a) An example explains how to calculate the intermediate vector representation $\vec{\mathbf{m}}_1$. For the sake of simplicity, we ignore the first term of Eqn. \eqref{eqn:vec_update}. The $f$-th vector in $\vec{\mathbf{m}}_{1}$ (i.e., $\vec{\mathbf{m}}_1^f$) can be interpreted as the force exerted by neighboring atom $j$ on atom $1$, where $w_{j1}^f$ is force magnitude and $\frac{\vec{\mathbf{r}}_{j1}}{\Vert \vec{\mathbf{r}}_{j1} \Vert}$ is the force direction. Subsequently, the total force exerted on atom $1$ is a linear combination of forces exerted on it by all other neighboring atoms $j$. Note that the forces are calculated $F$ times in parallel. (b) The update to $\vec{\mathbf{x}}_1^f$ is achieved through a linear combination of $F$ vectors within $\vec{\mathbf{m}}_1$. This computation is performed in parallel $F$ times to obtain $\vec{\mathbf{x}}_1$.}}
  \label{fgr:vector_visualization}
\end{figure}

In the $t$-th layer, during local message passing, a particular node $v_i$ gathers messages from its neighbouring scalar $\mathbf{x}_j^{(t)}$ and vector $\vec{\mathbf{x}}_j^{(t)}$, resulting in intermediate scalar and vector variables $\mathbf{m}_i$ and $\vec{\mathbf{m}}_i$ as follows:
\begin{equation}
    \mathbf{m}_i= \sum_{v_j \in \mathcal{N}(v_i)}( \mathbf{W}_h \mathbf{x}_{j}^{(t)}) \circ \lambda_h(\Vert \vec{\mathbf{r}}_{ji} \Vert)
\end{equation}
\begin{equation}
    \vec{\mathbf{m}}_i= \sum_{v_j \in \mathcal{N}(v_i)}(\mathbf{W}_u \mathbf{x}_{j}^{(t)}) \circ \lambda_u(\Vert \vec{\mathbf{r}}_{ji} \Vert) \circ \vec{\mathbf{x}}_j^{(t)} + (\mathbf{W}_v \mathbf{x}_{j}^{(t)}) \circ \lambda_v(\Vert \vec{\mathbf{r}}_{ji} \Vert) \circ \frac{\vec{\mathbf{r}}_{ji}}{\Vert \vec{\mathbf{r}}_{ji} \Vert}\label{eqn:vec_update}
\end{equation}
Here, $\mathbf{W}_h,\mathbf{W}_u,\mathbf{W}_v \in \mathbb{R}^{F \times F}$ are learnable matrices. The functions $\lambda_h$, $\lambda_u$, and $\lambda_v$ are the linear combination of Gaussian radial basis functions \cite{schutt2017schnet}. For the Eqn. \eqref{eqn:vec_update}, the first term $(\mathbf{W}_u \mathbf{x}_{j}^{(t)}) \circ \lambda_u(\Vert \vec{\mathbf{r}}_{ji} \Vert) \circ \vec{\mathbf{x}}_j^{(t)}$ propagates directional information $\vec{\mathbf{x}}_{j}^{(t)}$ obtained in the previous step to neighboring atoms, with $(\mathbf{W}_u \mathbf{x}_{j}^{(t)}) \circ \lambda_u(\Vert \vec{\mathbf{r}}_{ji} \Vert)$ acting as a gate signal to control how many signals in the previous step can be preserved. We interpret the second term as the force exerted by atom $j$ on atom $i$, where $(\mathbf{W}_v \mathbf{x}_{j}^{(t)}) \circ \lambda_v(\Vert \vec{\mathbf{r}}_{ji} \Vert)$ is force magnitude and $\frac{\vec{\mathbf{r}}_{ji}}{\Vert \vec{\mathbf{r}}_{ji} \Vert}$ is the force direction. This is conceptually different from PaiNN's message function, which uses $\vec{\mathbf{r}}_{ij}$ instead of $\vec{\mathbf{r}}_{ji}$. Subsequently, $\sum_{v_j \in \mathcal{N}(v_i)}(\mathbf{W}_v \mathbf{x}_{j}^{(t)}) \circ \lambda_v(\Vert \vec{\mathbf{r}}_{ji} \Vert) \circ \frac{\vec{\mathbf{r}}_{ji}}{\Vert \vec{\mathbf{r}}_{ji} \Vert}$ represents the total force exerted on atom $i$ and is a linear combination of forces exerted on it by all other atoms. \textbf{Fig. \ref{fgr:vector_visualization}a} shows an example of how the $F$ forces are calculated.

{\color{black}
\subsection{Local message updating}
The local message updating phase aims to aggregate $F$ scalars and vectors within $\mathbf{m}_i$ and $\vec{\mathbf{m}}_i$, respectively, to obtain new scalar $\mathbf{x}_i^{(t+1)}$ and new vector $\vec{\mathbf{x}}_i^{(t+1)}$, as shown in \textbf{Fig. \ref{fgr:vector_visualization}b}. Specifically, the scalar representation \( \mathbf{x}_i^{(t+1)} \) and vector representation \( \vec{\mathbf{x}}_i^{(t+1)} \) are updated according to the following equations:
\begin{equation}
    \mathbf{x}_i^{(t+1)}= \mathbf{W}_{s}(\mathbf{m}_{i} \oplus \Vert \mathbf{V}  \vec{\mathbf{m}}_i \Vert) + \tanh\big(\mathbf{W}_g(\mathbf{m}_{i} \oplus \Vert \mathbf{V}  \vec{\mathbf{m}}_i \Vert)\big) \mathbf{m}_{i}
\end{equation}
\begin{equation}
     \vec{\mathbf{x}}_i^{(t+1)}= \big( \mathbf{W}_{h}(\mathbf{m}_{i} \oplus \Vert \mathbf{V}  \vec{\mathbf{m}}_i \Vert) \big) \circ \big( \mathbf{U}\vec{\mathbf{m}}_i \big)
\end{equation}
where \( \oplus \) denotes concatenation, \( \mathbf{W}_{s}, \mathbf{W}_{g}, \mathbf{W}_{h} \in \mathbb{R}^{F \times 2F} \), and \( \mathbf{U}, \mathbf{V} \in \mathbb{R}^{F \times F} \). The term \( \tanh\big(\mathbf{W}_g(\mathbf{m}_{i} \oplus \Vert \mathbf{V} \vec{\mathbf{m}}_i \Vert)\big) \) acts as a gate controlling how much invariant signal from the previous layer is preserved. This design allows for selective updating of node representations, which can help in preserving important features from the previous layers while incorporating new information. 
}

{\color{black}
\subsection{Global message distributing and aggregating}
In many GNN architectures, information primarily flows locally between directly connected nodes. While increasing the depth of GNNs can broaden receptive fields, it can also lead to optimization instabilities, such as vanishing gradients and representation oversmoothing. To facilitate a more effective global communication channel across the entire graph, we propose a global message distributing and aggregating scheme by creating a global scalar \( \mathbf{x}_\mathcal{G} \) and a global vector \( \vec{\mathbf{x}}_\mathcal{G} \). Specifically, we initialize the global scalar and vector as \( \mathbf{x}_\mathcal{G}^{(0)} \in \mathbb{R}^F \) and \( \vec{\mathbf{x}}_\mathcal{G}^{(0)} = \vec{\mathbf{0}} \in \mathbb{R}^{F \times 3} \), where \( \mathbf{x}_\mathcal{G}^{(0)} \) is trainable. The global message distributing operates before the local message passing, distributing the global scalar and vector at the current step to each node using the following equations:
\begin{equation}
    \mathbf{x}_i^{(t)} = \phi(\mathbf{x}_i^{(t-1)} \oplus \mathbf{x}_\mathcal{G}^{(t-1)}) + \mathbf{x}_i^{(t-1)}
\end{equation}
\begin{equation}
    \vec{\mathbf{x}}_i^{(t)} = \mathbf{W}(\vec{\mathbf{x}}_i^{(t-1)} + \vec{\mathbf{x}}_\mathcal{G}^{(t-1)}) + \vec{\mathbf{x}}_i^{(t-1)}
\end{equation}
where $\phi: \mathbb{R}^{2F} \rightarrow \mathbb{R}^F$ refers to a multi-layer perceptron layer (MLP), and \( \mathbf{W} \in \mathbb{R}^{F \times F} \) is a trainable matrix. After local message updating, the global scalar and vector are updated using the node representations at the current step with the following equations:
\begin{equation}
    \mathbf{x}_\mathcal{G}^{(t+1)} = \phi\left(\left(\frac{1}{\vert \mathcal{G} \vert}\sum_{v_i \in \mathcal{G}} \mathbf{x}_i^{(t)}\right) \oplus \mathbf{x}_\mathcal{G}^{(t)}\right) + \mathbf{x}_\mathcal{G}^{(t)}
\end{equation}
\begin{equation}
    \vec{\mathbf{x}}_\mathcal{G}^{(t+1)} = \mathbf{W}\left(\left(\frac{1}{\vert \mathcal{G} \vert}\sum_{v_i \in \mathcal{G}} \vec{\mathbf{x}}_i^{(t)}\right) + \vec{\mathbf{x}}_\mathcal{G}^{(t)}\right) + \vec{\mathbf{x}}_\mathcal{G}^{(t)}
\end{equation}
LEIGNN is strictly equivariant to rotation and translation, as proven in \textbf{SM Section 1}. Its design also guarantees the extensivity of energy, i.e., the predicted energy of LEIGNN is linear with respect to the simulation box size, as proven in \textbf{SM Section 2}.
}

\subsection{Predicting potentials and forces}
To predict the potential \( e \), we utilize an MLP layer \( \phi: \mathbb{R}^F \rightarrow \mathbb{R}^1 \). This layer learns atom-wise potentials \( e_i \in \mathbb{R}^1 \) from the scalar representation \( \mathbf{x}_i^{(T)} \), which is obtained at the last graph convolution layer (referred to as the \( T \)-th layer). The total potential is then calculated as the sum of the atom-wise potentials:
\begin{equation}
e = \sum_{v_i \in \mathcal{G}} e_i
\end{equation}
The forces are predicted using the vector representation \( \vec{\mathbf{x}}_i^{(T)} \) as \( \vec{\mathbf{F}} = \mathbf{W}_f \vec{\mathbf{x}}_i^{(T)} \), where \( \mathbf{W}_f \in \mathbb{R}^{1 \times F} \). \textcolor{black}{This approach offers the advantage of creating a lightweight model by directly computing forces, which bypasses the computationally expensive tasks of calculating the gradient of the potential energy (first derivative) and the second derivative during backpropagation for model parameter updates. Although the directly predicted forces are not energy conserving, they can still facilitate MD simulations, especially when a thermostat is employed to regulate the temperature \cite{park2021accurate, fu2023forces}.}

\subsection{Implementation details}
The LEIGNN model is implemented using PyTorch. Experiments are conducted on an NVIDIA GeForce RTX A4000 with 16 GB of memory. The training objective aims to minimize the loss function defined as:
\begin{equation}
    \mathcal{L}= \frac{1}{N} \sum\limits_{n=1}^N \Big( \alpha \vert e_n-e_n^{l} \vert + \beta \frac{1}{3M} \sum\limits_{m=1}^M \sum\limits_{k=1}^3 \vert\vec{\mathbf{F}}_{nmk} -\vec{\mathbf{F}}_{nmk}^{l} \vert \Big)
\end{equation}
where $e_n^{l}$ represents the ground truth energy of $n$-th sample, $\vec{\mathbf{F}}_{nmk}^{l}$ is the ground truth force of $k$-th dimension of $m$-th atom in $n$-th sample. The variables $N$ and $M$ denote the sample size and the number of atoms in each sample, and $\alpha$ and $\beta$ denote the weights assigned to the energy and force losses, respectively.

\begin{table}[h]
\caption{Comparison results of the proposed LEIGNN and baselines on S2EF task of four external validation sets of OC20 in terms of energy MAE (meV) and forces MAE (meV/$\rm \AA$), where all models are trained on OC20-200K.}
\tabcolsep=0.18cm
\begin{tabular}{lllllllll}
\hline
\multirow{2}{*}{Model} & ID     &        & \multicolumn{2}{l}{OOD Ads.} & \multicolumn{2}{l}{OOD Cat.} & \multicolumn{2}{l}{OOD Both} \\ \cline{2-9} 
                       & Energy & Forces & Energy        & Forces       & Energy        & Forces       & Energy        & Forces       \\ \hline
CGCNN                  & 1111   & 75.0   & 1261          & 80.7         & 1097          & 74.1         & 1383          & 91.9         \\
SchNet                 & 975    & 59.6   & 1077          & 66.8         & 975           & 59.5         & 1204          & 77.9         \\
MACE & 565 & 51.3 & 657 & 59.8 & 589 & 51.3 & 802 & 68.6 
\\
PaiNN & 482 & 52.7 & 570 & 58.8 & 499 & 52.4 & 704 & 74.0     \\
PaiNN\_Direct & 457 & 46.6 & 571 & 53.5 & 497 & 46.9 & 689 & 63.4\\
DimeNet++              & 497    & 48.7   & 547           & 55.9         & 522           & 48.6         & 671           & 65.3         \\
GemNet-dT              & 443    & \textbf{41.3}   & 516           & \textbf{46.7}         & 548           & \textbf{41.6}         & 717           & \textbf{55.3}         \\
LEIGNN                 & \textbf{415}    & 43.9    &  \textbf{502}          &   49.6       & \textbf{452}           & 43.9         & \textbf{613}           & 58.1        \\ \hline
\label{tbl:OC20-200K}
\end{tabular}
\end{table}

\begin{figure}[htp]
  \centering
  \includegraphics[width=11.8cm]{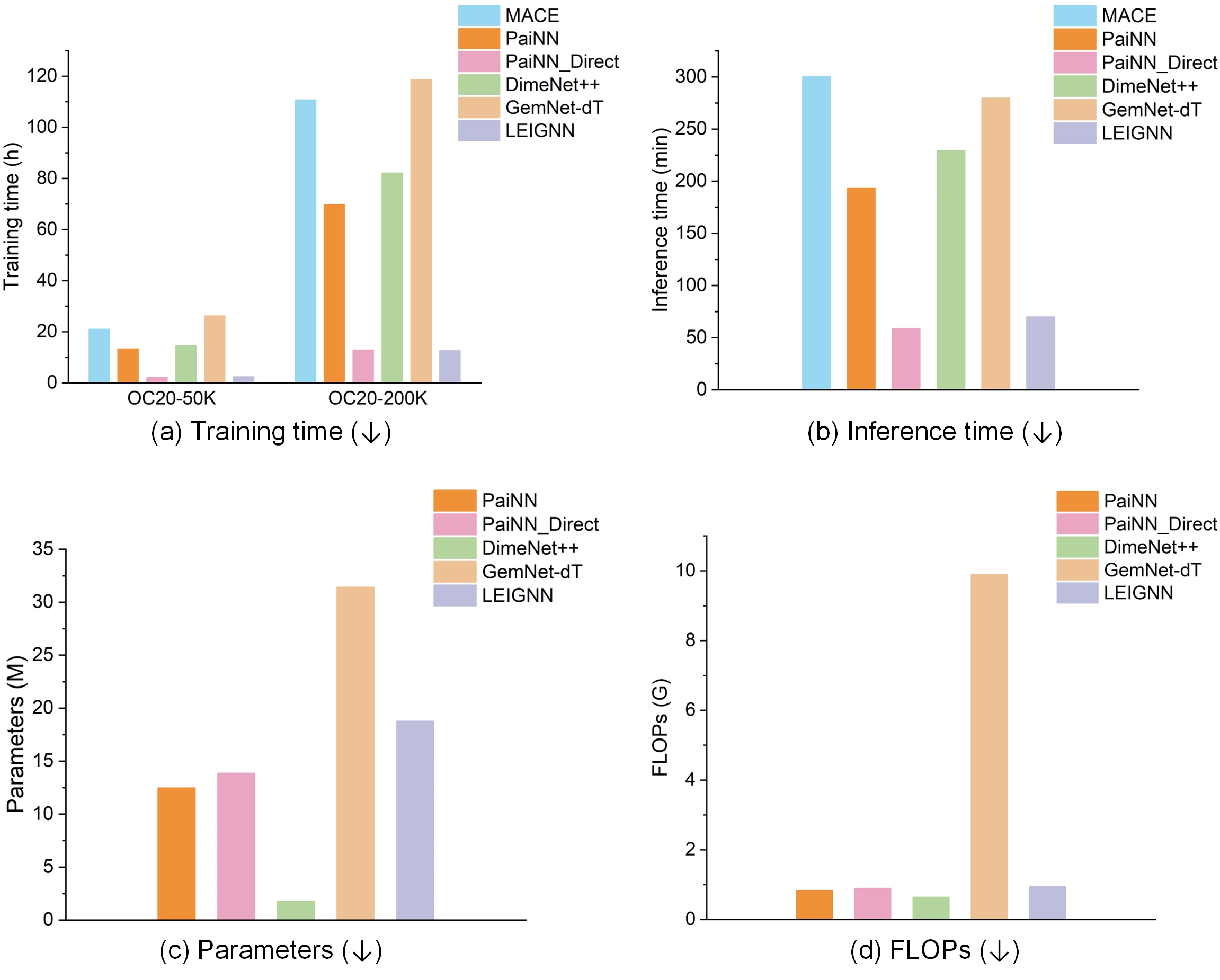}
  \caption{Computational costs of the representative models (the lower the better for all metrics). (a) Training time. (b) Mean inference time. (c) Number of parameters. (d) FLOPs.}
  \label{fgr:complexity}
\end{figure}

\section{Results}
\subsection{Model Performance}

We first evaluate LEIGNN in the structure to energy and forces (S2EF) task using Open Catalyst 2020 (OC20) dataset \cite{chanussot2021open}. The purpose of this task is to predict energies and forces corresponding to each trajectory during structural relaxation. The OC20 dataset encompasses 1,281,040 density functional theory (DFT) relaxations with 264,890,000 single-point calculations and spans a vast range of materials, surfaces, and adsorbates. It has been reported that training a GNN model on the whole OC20 dataset requires hundreds or even thousands of days \cite{passaro2023reducing}. Therefore, we only use a subset of it: OC20-200K ($N=200,000$). The dataset is split into a training set and an internal validation set with a ratio of 8:2, where the internal validation set is used to select the best model for testing. Finally, the select models are tested on four external validation sets provided by the OC20 project: in Domain (ID), out-of-domain adsorbate (OOD Ads.), out-of-domain catalyst (OOD Cat.), and OOD Both (both the adsorbate and catalyst are not seen in the training set). Each external validation set contains approximately 1M data points. \textcolor{black}{We compare LEIGNN with seven representative baseline models: CGCNN \cite{xie2018crystal}, SchNet \cite{schutt2017schnet}, MACE \cite{batatia2022mace}, PaiNN \cite{schutt2021equivariant}, PaiNN\_Direct \cite{tran2023open}, DimeNet++ \cite{gasteiger_dimenetpp_2020}, and GemNet-dT \cite{gasteiger2021gemnet}.} PaiNN\_Direct is a modified version of PaiNN designed for making direct force predictions. The hyperparameter configuration for each model is provided in \textbf{SM Section 3}. All models share the same training, internal validation, and external validation sets, and are trained to predict adsorption energy and per-atom forces simultaneously. The performance of the models is evaluated based on the mean absolute error (MAE).

{\color{black}From \textbf{Table \ref{tbl:OC20-200K}}, we have two key observations. First, LEIGNN achieves performance close to GemNet-dT and outperforms other baseline models. Second, PaiNN\_Direct surpasses PaiNN, which aligns with previously reported results \cite{kolluru2022open}. Additional experimental results on a smaller subset, OC20-50K, are available in \textbf{SM Section 4}.

Next, we compare the computational costs among LEIGNN and five accurate models: MACE, PaiNN, PaiNN\_Direct, DimeNet++, and GemNet-dT. \textbf{Fig. \ref{fgr:complexity}} illustrates the model complexity in terms of training time, inference time, the number of parameters, and floating-point operations (FLOPs). It is worth noting that the number of parameters and FLOPs are estimated using PyTorch-OpCounter, which fails when applied to MACE. We have three key observations. First, LEIGNN offers approximately ten times faster training and inference speeds than GemNet-dT and MACE, and about five times faster than DimeNet++ and PaiNN. Second, LEIGNN has a similar computational cost to PaiNN\_Direct but demonstrates better performance. Third, the computational cost of PaiNN\_Direct is significantly lower than that of PaiNN. Considering that the main difference between PaiNN\_Direct and PaiNN is how they compute forces, we can conclude that computing forces by taking the derivative of the energy with respect to positions is more computationally demanding. Overall, our results show that direct force prediction improves both efficiency and performance, consistent with previous studies \cite{kolluru2022open, shuaibi2021rotation, hu2021forcenet}. Besides the OC20 database, we compare LEIGNN with other models on MD17 and ISO17 databases. Our LEIGNN model also surpasses those benchmark models on these two databases (see \textbf{SM Sections 7 and 8}), showing the universality of LEIGNN.}

\begin{table}[h]
\caption{Ablation study on the OC20-50K and OC20-200K datasets to demonstrate the effectiveness of LEIGNN. The improvement is calculated by comparing LEIGNN with the Vanilla model.}
\tabcolsep=0.08cm
\begin{tabular}{llllllllll}
\hline
\multirow{2}{*}{Dataset}   & \multirow{2}{*}{Models} & \multicolumn{2}{l}{ID} & \multicolumn{2}{l}{OOD Ads.} & \multicolumn{2}{l}{OOD Cat.} & \multicolumn{2}{l}{OOD Both} \\ \cline{3-10} 
                           &                         & Energy     & Forces    & Energy        & Forces       & Energy        & Forces       & Energy        & Forces       \\ \hline
\multirow{4}{*}{OC20-50K}  & Vanilla             & 568        & 53.1      & 691           & 57.9         & 602           & 52.6         & 847          & 68.3         \\
& Vanilla + NMU  & 552 & 52.4 & 634 & 57.4 & 615 & 51.8 & 785 & 67.7 \\
& LEIGNN          & \textbf{531}        & \textbf{51.2}      & \textbf{626}           & \textbf{56.2}         & \textbf{555}           & \textbf{50.7}         & \textbf{750}           & \textbf{66.0}        \\    
                           & Improvement & 6.52\% & 3.58\% & 9.40\% & 2.93\% & 7.81\% & 3.61\% & 11.45\% & 3.37\%   \\ \hline
\multirow{4}{*}{OC20-200K} & Vanilla & 434 & 44.7 & 519 & 50.7 & 487 & 45.0 & 659 & 59.7 \\
& Vanilla + NMU  & 427 & 44.7 & 525 & 50.6 & 480 & 44.8 & 662 & 59.4       \\
& LEIGNN                 & \textbf{415}    & \textbf{43.9}    &  \textbf{502}          &   \textbf{49.6}       & \textbf{452}           & \textbf{43.9}         & \textbf{613}           & \textbf{58.1} \\
                          & Improvement & 4.38\% & 1.79\% & 3.28\% & 2.17\% & 7.18\% & 2.44\% & 6.98\% & 2.68\%     \\ \hline
\label{tbl:Ablation}
\end{tabular}
\end{table}

{\color{black}
\subsection{Ablation study}
The novelty of LEIGNN lies in two aspects: a novel message updating (NMU) scheme and the global message aggregating and distributing phases to promote long-range information exchange among atoms. To demonstrate the effectiveness of these strategies, we compare LEIGNN with two baseline models: Vanilla and Vanilla + NMU. The Vanilla model excludes both the NMU scheme and the global message phases, using PaiNN's message updating instead. The Vanilla + NMU model includes only the NMU scheme. As shown in Table \ref{tbl:Ablation}, LEIGNN notably outperforms both baseline models. Besides, it is worth noting that such two strategies are carried out on nodes rather than edges. This design choice offers a computational advantage because the number of nodes is typically significantly smaller than the number of edges in a graph.}

\begin{figure}[htbp]
  \centering
  \includegraphics[width=11.8cm]{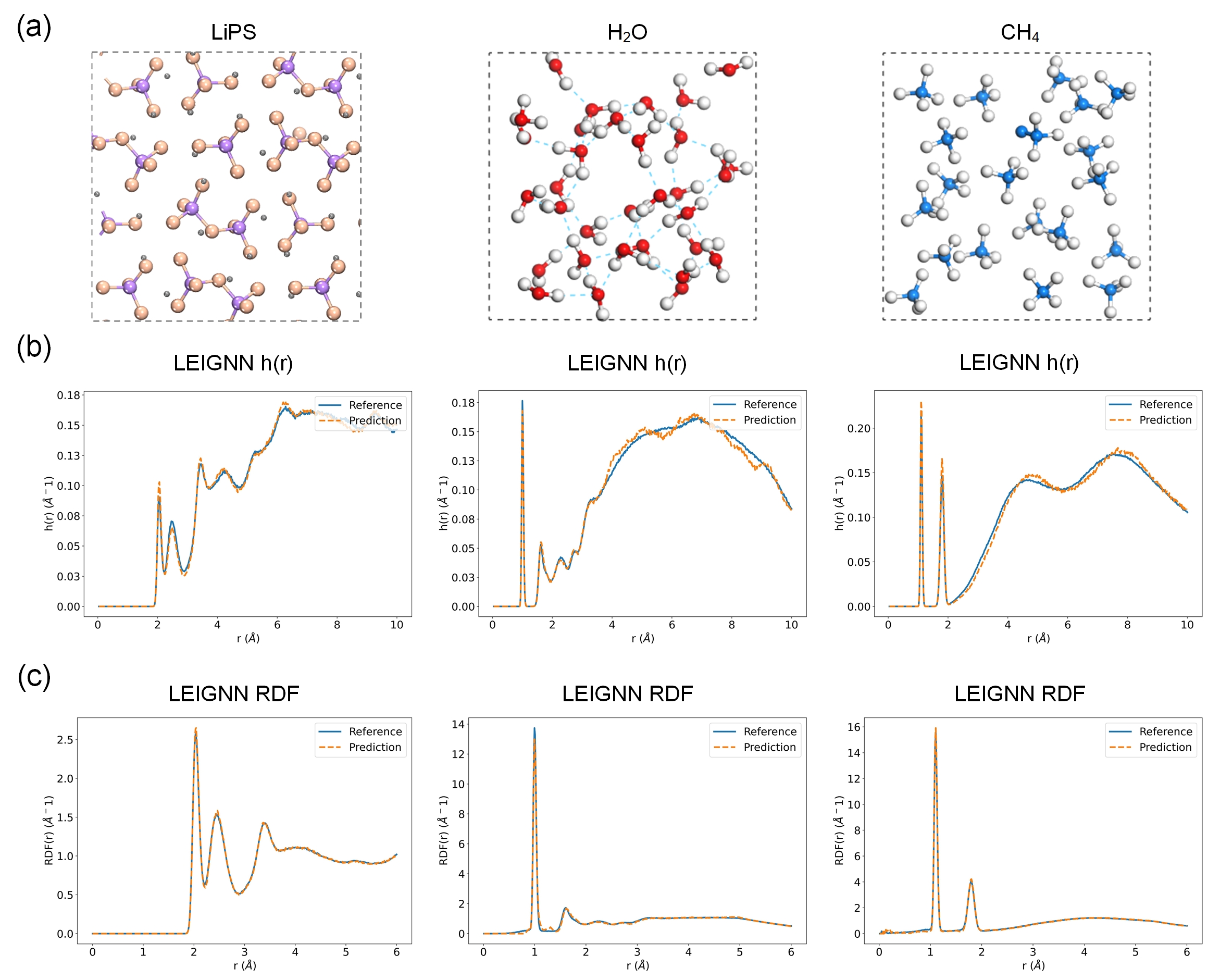}
  \caption{MD simulations for LiPS, $\rm H_2O$, and $\rm CH_4$. (a) An overview of the three benchmark systems. (b) $h(r)$ for trajectories predicted by the LEIGNN. (c) RDF for trajectories predicted by LEIGNN.}
  \label{fgr:MD_crystal}
\end{figure}

\begin{table}[h]
{\color{black}
\caption{Evaluating LEIGNN and the representative models on MD simulations using LiPS system in terms of forces, stability (ps), MAE of RDF  (unitless), and diffusivity ($10^{-9} \ \rm m^2/s$).} \label{tbl:MD_LiPS}
\centering
\tabcolsep=0.35cm
\begin{tabular}{lllll}
\hline
Model      & Force (↓) & Stability (↓) & RDF (↓) & Diffusivity (↓) \\ \hline
LJ potential & - & 0 & 12.34 & - \\
DeepPot-SE$^*$ & 40.5      & 4         & 0.27    & -               \\
SchNet$^*$     & 28.8      & 50        & 0.04    & 0.38            \\
DimeNet$^*$    & 3.2       & 48        & 0.05    & 0.30            \\
ForceNet$^*$  & 12.8      & 26        & 0.51    & -               \\
GemNet-T$^*$   & 1.3       & 50        & 0.04    & 0.24            \\
GemNet-dT$^*$  & 1.4       & 50        & 0.04    & 0.28            \\
NequIP$^*$     & 3.7       & 50        & 0.04    & 0.34            \\
LEIGNN     & 1.5       & 50        & 0.05    & 0.21            \\ \hline
\end{tabular}
\\
\begin{flushleft}
\footnotesize{$^*$These results are taken from \cite{fu2023forces}. LEIGNN is evaluated under the same conditions as the reported results.}
\end{flushleft}}
\end{table}

\subsection{Molecular dynamics simulations}
Molecular dynamics simulations provide vital atomistic insights, but the accuracy versus efficiency trade-off has long been a challenge. Classical MD, which relies on empirical interatomic potentials, is computationally efficient but sacrifices accuracy. In contrast, AIMD, which integrates first-principles methods such as density functional theory, provides higher accuracy but at a high computational cost. Recently, machine learning-based force fields have emerged as a promising solution to enhance the speed of MD simulations by several orders of magnitude while maintaining quantum chemical accuracy.

{\color{black}To validate the interatomic potentials of our LEIGNN model, we demonstrate that LEIGNN performs competitively against established baseline models such as Lennard-Jones (LJ) potential, DeepPot-SE \cite{zhang2018end}, SchNet \cite{schutt2017schnet}, DimeNet \cite{gasteiger_dimenet_2020}, ForceNet \cite{hu2021forcenet}, GemNet-T \cite{gasteiger2021gemnet}, GemNet-dT \cite{gasteiger2021gemnet}, and NequIP \cite{batzner20223}. We evaluate the models' performance using the mean absolute error (MAE) of the radial distribution function (RDF), stability, and diffusivity. Our definitions of stability and diffusivity align with those used in MDSim \cite{fu2023forces}. A simulation is considered ``unstable" if deviations exceed predefined thresholds, leading to the sampling of highly nonphysical structures. Diffusivity measures the time-correlation of translational displacement. For detailed experimental settings and metrics, refer to \textbf{SM Sections 5 and 6}.

We consider three systems: LiPS (solid), $\rm H_2O$ (liquid), and $\rm CH_4$ (gas), as depicted in \textbf{Fig. \ref{fgr:MD_crystal}a}, where the data for $\rm H_2O$ and $\rm CH_4$ (gas) are collected in-house to assess LEIGNN's effectiveness in liquid and gaseous systems. We conduct a 50-picosecond simulation for each system, starting from a randomly selected test configuration. The training, validation, and testing sets for the LiPS system are consistent with those used in MDSim. The results for the LiPS system are summarized in \textbf{Table \ref{tbl:MD_LiPS}}. Although the forces predicted by LEIGNN are not from the widely used energy-conserving method, the MD simulation performance remains competitive with that of energy-conserving models. \textbf{Fig. \ref{fgr:MD_crystal}b-c} show the $h(r)$ and RDF for AIMD and LEIGNN across the three systems, revealing their strong alignment. Using the same stability criterion as for the LiPS system, LEIGNN also demonstrates consistent stability throughout the 50 ps simulations in the $\rm H_2O$ (liquid) and $\rm CH_4$ (gas) systems. These results highlight LEIGNN's potential as a crucial tool in computational materials science.}

\section{Discussion}
\textcolor{black}{We develop a lightweight, accurate, and efficient equivariant graph neural network for predicting interatomic potentials and forces. The lightweight nature of LEIGNN comes from two factors: direct force predictions and modeling only two-body interactions. LEIGNN consistently outperforms the prediction performance of the representative baselines across diverse molecular and crystal datasets. Moreover, LEIGNN achieves the precision of $ab~initio$ MD while maintaining high computational efficiency in simulations involving gas, liquid, and solid systems. Despite its advancements, we would like to say two limitations. First, the introduction of global entities may lead to inaccuracies in certain scenarios where distant parts of the system are incorrectly influenced by global changes, such as two ``isolated" molecules in a big box without interaction between them. Second, the forces are predicted directly, which is not energy conserving and a thermostat is needed to regulate the temperature \cite{fu2023forces, hu2021forcenet}.}

\section*{Data availability}
The data that support the findings of this study are available in \url{https://github.com/Shen-Group/LEIGNN}.

\section*{Code availability}
The code for LEIGNN, along with detailed instructions, is available at  \url{https://github.com/Shen-Group/LEIGNN}. 

\section*{Acknowledgements}
This work was supported by the National Natural Science Foundation of China (Grant No. 62176272), Research and Development Program of Guangzhou Science and Technology Bureau (No. 2023B01J1016), Key-Area Research and Development Program of Guangdong Province (No. 2020B1111100001), Singapore MOE Tier 1 (No. A-8001194-00-00), and Singapore MOE Tier 2 (No. A-8001872-00-00).

\section*{Author contributions}
Lei Shen and Ziduo Yang conceptualized and designed the research. Ziduo Yang and Xian Wang conducted the experiments. Ziduo Yang, Xian Wang, Yifan Li, Qiujie Lv, Calvin Yu-Chian Chen, and Lei Shen analyzed the data and interpreted the results. Ziduo Yang and Lei Shen jointly wrote the manuscript.

\section*{Competing interests}
The author declares no competing interests.


\bibliography{sn-bibliography}


\end{document}